\documentclass[journal]{IEEEtran}
\usepackage{mathrsfs}
\usepackage{pifont}
\usepackage{bbding}
\usepackage{tipa}
\usepackage{amsmath,epsfig}
\usepackage{stmaryrd}
\usepackage{amssymb}
\usepackage{amsfonts}
\usepackage{epic}
\usepackage{graphicx}
\usepackage{curves}
\ifCLASSINFOpdf
\else
\fi
\begin{document}
\newtheorem{theorem}{Theorem}
\newtheorem{proposition}{Proposition}
\newtheorem{definition}{Definition}
\newtheorem{lemma}{Lemma}
\newtheorem{corollary}{Corollary}
\newtheorem{remark}{Remark}
\newtheorem{construction}{Construction}

\newcommand{\supp}{\mathop{\rm supp}}
\newcommand{\sinc}{\mathop{\rm sinc}}
\newcommand{\spann}{\mathop{\rm span}}
\newcommand{\essinf}{\mathop{\rm ess\,inf}}
\newcommand{\esssup}{\mathop{\rm ess\,sup}}
\newcommand{\Lip}{\rm Lip}
\newcommand{\sign}{\mathop{\rm sign}}
\newcommand{\osc}{\mathop{\rm osc}}
\newcommand{\R}{{\mathbb{R}}}
\newcommand{\Z}{{\mathbb{Z}}}
\newcommand{\C}{{\mathbb{C}}}
%
% paper title
% can use linebreaks \\ within to get better formatting as desired

%=======================================================title==============================================================================
\title{Joint Power Allocation and Precoding for Network Coding based Cooperative Multicast Systems}
% author names and affiliations
% use a multiple column layout for up to three different
% affiliations
%=======================================================author information=================================================================
\author{{Jun~Li, and}
        Wen~Chen,~\IEEEmembership{Member,~IEEE}
        % <-this % stops a space
\thanks{Jun~Li and Wen~Chen are with the Department of Electronic Engineering, Shanghai Jiaotong University, Shanghai, 200240
PRC. e-mail: \{jleesr80, wenchen\}@sjtu.edu.cn}.% <-this % stops a space
\thanks{This work is supported by NSF China \#60672067, by NSF Shanghai
\#06ZR14041, by Shanghai-Canada NRC \#06SN07112, by Cultivation Fund
of the Key Scientific and Technical Innovation Project, Ministry of
Education of China \#706022, by Program for New Century Excellent
Talents in University \#NCET-06-0386, by PUJIANG Talents \#07PJ4046,
and by Huawei Fund for Sciences and Technologies in Universities
\#YJCB2008048WL}}

% conference papers do not typically use \thanks and this command
% is locked out in conference mode. If really needed, such as for
% the acknowledgment of grants, issue a \IEEEoverridecommandlockouts
% after \documentclass

% for over three affiliations, or if they all won't fit within the width
% of the page, use this alternative format:
%
%\author{\IEEEauthorblockN{Michael Shell\IEEEauthorrefmark{1},
%Homer Simpson\IEEEauthorrefmark{2},
%James Kirk\IEEEauthorrefmark{3},
%Montgomery Scott\IEEEauthorrefmark{3} and
%Eldon Tyrell\IEEEauthorrefmark{4}}
%\IEEEauthorblockA{\IEEEauthorrefmark{1}School of Electrical and Computer Engineering\\
%Georgia Institute of Technology,
%Atlanta, Georgia 30332--0250\\ Email: see http://www.michaelshell.org/contact.html}
%\IEEEauthorblockA{\IEEEauthorrefmark{2}Twentieth Century Fox, Springfield, USA\\
%Email: homer@thesimpsons.com}
%\IEEEauthorblockA{\IEEEauthorrefmark{3}Starfleet Academy, San Francisco, California 96678-2391\\
%Telephone: (800) 555--1212, Fax: (888) 555--1212}
%\IEEEauthorblockA{\IEEEauthorrefmark{4}Tyrell Inc., 123 Replicant Street, Los Angeles, California 90210--4321}}

% use for special paper notices
%\IEEEspecialpapernotice{(Invited Paper)}
\markboth{IEEE Signal Processing Letters (Submitted) }{Shell
\MakeLowercase{\textit{et al.}}: Bare Demo of IEEEtran.cls for
Journals}
% make the title area
\maketitle

%=======================================================abstract=================================================================
\begin{abstract}
%\boldmath
In this letter, we propose two power allocation schemes based on the
statistical channel state information (CSI) and instantaneous
$s\rightarrow{r}$ CSI at transmitters respectively for a $2-N-2$
cooperative multicast system with non-regenerative network coding.
Then the isolated precoder and the distributed precoder are
respectively applied to the schemes to further improve the system
performance by achieving the full diversity gain. Finally, we
demonstrate that joint instantaneous $s\rightarrow{r}$ CSI based
power allocation and distributed precoder design achieve the best
performance.
\end{abstract}

% IEEEtran.cls defaults to using math in the Abstract.
% This preserves the distinction between vectors and scalars. However,
% if the conference you are submitting to favors bold math in the abstract,
% then you can use LaTeX's standard command \boldmath at the very start
% of the abstract to achieve this. Many IEEE journals/conferences frown on
% math in the abstract anyway.

% no keywords

% For peer review papers, you can put extra information on the cover
% page as needed:
% \ifCLASSOPTIONpeerreview
% \begin{center} \bfseries EDICS Category: 3-BBND \end{center}
% \fi
%
% For peerreview papers, this IEEEtran command inserts a page break and
% creates the second title. It will be ignored for other modes.
\IEEEpeerreviewmaketitle

%=======================================================section1 introduction=======================================================
\section{Introduction}\label{sec:1}
% no \IEEEPARstart
Network coding has been proved to achieve the network multicast
capacity bound in the wireline systems \cite{IEEEconf:1}. Recently,
how to leverage network coding in wireless physical layer networks
for system capacity improvement has drawn increasing interest
\cite{IEEEconf:2}-\cite{IEEEconf:4}. However, these works are based
on the multi-access relay channels model and unicast model.

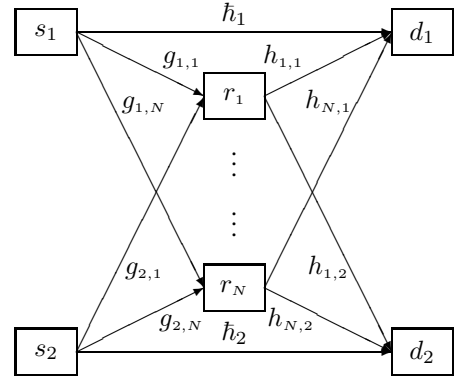
\begin{figure}[!t]
\begin{picture}(80,50)(-6,0)
\put(10,43){\framebox(8,6){$s_1$}}
%====================================================================
\put(60,43){\framebox(8,6){$d_1$}} \put(18,46){\vector(1,0){42}}
\put(35,47){\makebox(8,6)[b]{$\hbar_1$}}
%====================================================================
\put(35,34.5){\framebox(8,6){$r_{\scriptscriptstyle{1}}$}}
\put(18,46){\vector(2,-1){17}} \put(43,37.5){\vector(2,1){17}}
\put(32,41){\makebox(0,0)[b]{$g_{\scriptscriptstyle{1,1}}$}}
\put(44,41){\makebox(3,2)[b]{$h_{\scriptscriptstyle{1,1}}$}}
\put(32,6){\makebox(0,0)[b]{$g_{\scriptscriptstyle{2,N}}$}}
\put(45,6){\makebox(3,2)[b]{$h_{\scriptscriptstyle{N,2}}$}}
\put(43,12){\vector(2,-1){17}} \put(18,3.5){\vector(2,1){17}}
%====================================================================
\put(35,18.5){\makebox(8,6){$\vdots$}}
\put(35,26.5){\makebox(8,6){$\vdots$}}
%====================================================================
\put(35,9){\framebox(8,6){$r_{\scriptscriptstyle{N}}$}}
\put(18,46){\vector(1,-2){17}} \put(43,12){\vector(1,2){17}}
\put(27,35){\makebox(0,0)[b]{$g_{\scriptscriptstyle{1,N}}$}}
\put(50,35){\makebox(3,2)[b]{$h_{\scriptscriptstyle{N,1}}$}}
\put(27,13){\makebox(0,0)[b]{$g_{\scriptscriptstyle{2,1}}$}}
\put(50,13){\makebox(3,2)[b]{$h_{\scriptscriptstyle{1,2}}$}}
\put(43,37.5){\vector(1,-2){17}} \put(18,3.5){\vector(1,2){17}}
%====================================================================
\put(10,0.5){\framebox(8,6){$s_2$}}\put(60,0.5){\framebox(8,6){$d_2$}}
\put(18,3.5){\vector(1,0){42}}
\put(35,4.5){\makebox(8,6)[b]{$\hbar_2$}}
\end{picture} \caption{Multicast cooperative channels with $N$ relays.} \label{fig1}
\end{figure}

In this letter, we study the multicast model with two sources, two
destinations and $N$ relays ($2-N-2$ system) by following the second
scheduling strategy in~\cite{IEEEconf:5}, where relays are arranged
in the round-robin way. We suppose that $s_1$ as well as $s_2$
broadcast their information to the two destinations $d_1$ and $d_2$
simultaneously. From Fig.~\ref{fig1}, we can see $d_1$ (or $d_2$) is
out of the transmission range of $s_2$ (or $s_1$). The shared relays
can help $s_1$ (or $s_2$) reach their destinations. By the wireless
network coding method, there are two time slots:
\\\indent 1. $s_1\rightarrow\{r,~d_1\}$ with $X_{s_1}$; $s_2\rightarrow\{r,~d_2\}$ with $X_{s_2}$,
\\\indent 2. $r\rightarrow\{d_1,~d_2\}$ with $f(X_{s_1},X_{s_2})$,
\\where $f(\cdot)$ is certain mapping mechanism. We focus on the
non-regenerative network coding where mixed signals are scaled and
retransmitted to the destinations without decoding at relays. We
propose two power allocation schemes based on the statistical
channel state information (CSI) and the instantaneous
$s\rightarrow{r}$ CSI at sources respectively. The isolated
precoders and distributed precoders are respectively jointly working
with the schemes to achieve the full diversity gain. We will
demonstrate that the joint instantaneous $s\rightarrow{r}$ CSI based
power allocation and distributed precoder can achieve the best
performance against the other combinations in the terms of system
frame error probability (SFEP).
% You must have at least 2 lines in the paragraph with the drop letter
% (should never be an issue)
%=======================================================section2=====================================================================
\section{System Model}\label{sec:2}
Channel coefficients are shown in Fig.~\ref{fig1} with zero means
and unit variances. The noise variances are equal to $\sigma^2$ in
all the receivers. To normalize the power, we denote $P$ as the
average total network transmission power over a time slot. Then we
define the system SNR as $\rho\triangleq\frac{P}{\sigma^2}$. We
define $\textbf{x}_{s}\triangleq
[x_{s_1,1},x_{s_2,1},x_{s_1,2},x_{s_2,2},\cdots,x_{s_1,N},x_{s_2,N}]^T$
as a system frame which is composed of symbols from two sources. All
symbols are equally probable from the QAM constellation set with
zero means and variances $2P$. The symbol vector in the system frame
for $s_k\,(k=1,\,2)$ is denoted as
$\textbf{x}_{s_k}=[x_{s_k,1},x_{s_k,2},\cdots,x_{s_k,N}]^T$. Since
each relay is only used once during a frame period, we denote the
signal received by the $i$-th relay as $y_{r_i}$. The symbol vector
received by the $k$-th destination is denoted as
$\textbf{y}_{d_k}=[y_{d_k,1},y_{d_k,2},\cdots,y_{d_k,2N-1},y_{d_k,2N}]^T$.

The total power consumed during a frame period is
\begin{equation}\label{equ:1}
\mathcal{E}\left\{\textbf{k}_1^T|\textbf{x}_{s_1}|^2+\textbf{k}_2^T|\textbf{x}_{s_2}|^2+|\textbf{b}^T\textbf{y}_r|^2\right\}=2NP
\end{equation}
where $\textbf{y}_{r}=[y_{r_1},y_{r_2},\cdots,y_{r_N}]^T$. The power
is distributed according to the power allocation factor vectors,
i.e., $\textbf{k}_k=[\kappa_{k,1},\cdots,\kappa_{k,N}]^T$ is the
factor vector for the $k$-th source and
$\textbf{b}=[b_1,\cdots,b_N]^T$ is the amplification factor vector
where
$b_i=\sqrt{\frac{2\tau_i{P}}{2\kappa_{1,i}{P}+2\kappa_{2,i}{P}+\sigma^2}}$
is the factor for the $i$-th relay. Note that $\tau_i$ is the
corresponding power allocation factor for the $i$-th relay. We
denote $\textbf{t}=[\tau_1,\cdots,\tau_N]^T$.

The received signals of the $k$-th destination for the $(2i-1)$-th
and $2i$-th time slots are respectively written as
\begin{equation}\label{equ:2}
y_{d_k,2i-1}=\hbar_k\sqrt{\kappa_{k,i}}x_{s_k,i}+v_{d_k,2i-1}\text{,}\qquad\qquad\qquad\qquad\;\;
\end{equation}
\begin{equation*}
\begin{split}
y_{d_k,2i}=b_ih_{i,k}(g_{1,i}\sqrt{\kappa_{1,i}}x_{s_1,i}+g_{2,i}\sqrt{\kappa_{2,i}}x_{s_2,i}+v_{r_{i}})+v_{d_k,2i}.
\end{split}
\end{equation*}
Note that $v_{r_i}$ is the noise observed by the $i$-th relay and
$v_{d_k,2i}$ is the noise observed by $d_k$ in the $2i$-th time
slot. Joint ML decoding is performed at the end of each frame
period.
%=======================================================section3=====================================================================
\section{Performance Analysis and Improvements}\label{sec:3}
We measure the performance by system frame error probability (SFEP).
We define that a frame is successfully transmitted if and only if
both destinations can successfully receive the frame. So the SFEP of
the multicast system is
\begin{equation}\label{equ:3}
P_{sys}=P_{d_k}(1-P_{d_{\bar{k}}})+P_{d_{\bar{k}}}(1-P_{d_k})+P_{d_k}{P}_{d_{\bar{k}}}
\end{equation}
where $\bar{k}$ is the complementary element of $k$ in set
$\{1,2\}$, and $P_{d_k}$ is the FEP of $d_k$. We improve the
performance by combining power allocation and precoders according to
the knowledge of the CSI at sources.
%=====================================================subsection A=============================================
\subsection{Statistical CSI based Power Allocation}
Due to the symmetrical quality of $P_{sys}$ and according to the
fact that each channel variance is equal, power should be equally
allocated to each symbol of the two sources, i.e.,
$\textbf{k}_1=\textbf{k}_2=[\kappa,\cdots,\kappa]^T$ and
$\textbf{t}=[\tau,\cdots,\tau]^T$. Then the amplification factor
$b=\sqrt{\frac{2\tau{P}}{4\kappa{P}+\sigma^2}}$. Then we get the
statistical CSI based power allocation scheme as follows.
\begin{theorem}\label{theo1}
When $\rho$ is large enough, the statistical CSI based optimal power
allocation scheme chooses the power allocation factors as
$\kappa=\frac1{2(N+1)}$ and
$\tau=\frac{N}{N+1}$.~~~~~~~~~~~~~{\ding{111}}
\end{theorem}

\emph{Proof:} On the condition that only statistical CSI is
available, $P_{d_k}$ can be deduced by the average pairwise error
probability (PEP). Note that $\text{FEP}=2^{2RN}\text{PEP}$ since
there are total $2^{2RN}$ codewords, where $R$ is the transmission
rate. So to find out the optimal relation between $\kappa$ and
$\tau$, we first come to the average PEP of the destinations. We
rewrite (\ref{equ:2}) in matrix form as $\textbf{x}_{d_k}=
\textbf{X}_{2N}\textbf{h}_{2N} + \textbf{v}_{2N}$ and get the PEP by
developing the method in \cite{IEEEconf:7}.
\begin{equation}
P_{PE,d_k}=\qquad\qquad\qquad\qquad\qquad\qquad\qquad\qquad\qquad
\end{equation}
\begin{equation*}\label{equ:4}
\qquad\frac1{\pi}\int_0^{\frac{\pi}{2}}\mathcal
{E}_{\textbf{h}_{2N}}\left\{\exp\left(-\rho\frac{\textbf{h}_{2N}^H\textbf{U}_{2N}^H{\boldsymbol{\Sigma}}_{\textbf{v}}^{-1}\textbf{U}_{2N}\textbf{h}_{2N}}{8\sin^2\theta}\right)\right\}\,\mathrm{d}{\theta},
\end{equation*}
where $\textbf{U}_{2N}=\textbf{X}_{2N}-\hat{\textbf{X}}_{2N}$ is the
decoding error matrix and ${\boldsymbol{\Sigma}}_{\textbf{v}}$ is
the variance matrix of $\textbf{v}_{2N}$. $\textbf{h}_{2N}$ can be
written by $\textbf{T}_{2N}\textbf{g}_{2N}$ where
$\textbf{g}_{2N}=[\hbar_k,g_{1,1},g_{2,1},g_{1,2},g_{2,2},\cdots,g_{1,N},g_{2,N}]^T$
and
$\textbf{T}_{2N}=\mathrm{diag}(1,bh_{1,k},bh_{1,k},bh_{2,k},bh_{2,k},\cdots,bh_{N,k},bh_{N,k})$.
Note that for a random column vector
$\textbf{z}\sim\mathcal{N}(0,\boldsymbol\Sigma_{\textbf{z}})$ and a
Hermitian matrix $\textbf{H}$, there is
$\mathcal{E}[\exp(-\textbf{z}^H\textbf{Hz})]=1/\det(\textbf{I}+\boldsymbol{\Sigma_{\textbf{z}}}\textbf{H})$.
We take expectation with respect to $\textbf{g}_{2N}$ and let
$\textbf{y}=[y_1,y_2,\cdots,y_N]$ where $y_i=|h_{i,k}|^2$ with the
probability distribution function $e^{-{y_{i}}}$. Then the PEP is
written as
\begin{equation}\label{equ:5}
P_{PE,d_k}=\frac1{\pi}\int_0^{\frac{\pi}{2}}\int_0^{\infty}\cdots\int_0^{\infty}\frac{\exp\left(-\overset{N}{\underset{j=1}{\boldsymbol{\sum}}}{y_j}\right)}{\det(\textbf{A})}\,\mathrm{d}{\textbf{y}}\,\mathrm{d}{\theta}.
\end{equation}
In (\ref{equ:5}),
$\det(\textbf{A})=\det(\textbf{A}_0)\det(\textbf{A}_1)\cdots\det(\textbf{A}_N)$,
where
\begin{equation}\label{equ:6}
\begin{split}
&\textbf{A}_0=1+\frac{\rho}{8\sin^2\theta}\overset{N}{\underset{j=1}{\boldsymbol\sum}}|u_{s_k,j}|^2,
\\&\textbf{A}_i=
\begin{pmatrix}
1+\lambda_i|u_{s_1,i}|^2 &
\lambda_{i}u_{s_1,i}^{*}u_{s_2,i}\\\lambda_{i}u_{s_1,i}u_{s_2,i}^* &
1+\lambda_i|u_{s_2,i}|^2
\end{pmatrix},
\end{split}
\end{equation}
$u_{s_k,j}=\sqrt{{\kappa}/{P}}(x_{s_k,j}-\hat{x}_{s_k,j})$ is the
decoding error value of the $j$-th symbol, and
$\lambda_{i}=\frac{b^2|h_{i,k}|^2\rho}{8\sin^2{\theta}(1+b^2|h_{i,k}|^2)}$.
So
$\det(\textbf{A}_i)=1+\lambda_{i}|u_{s_1,i}|^2+\lambda_{i}|u_{s_2,i}|^2$.
Then we turn to the integral for
$\int_0^{\infty}\frac{e^{-{y_i}}}{\det(\textbf{A}_i)}\,\mathrm{d}{y_i}$,
i.e.,
\begin{equation}\label{equ:7}
\begin{split}
&\int_0^{\infty}\frac{e^{-{y_i}}}{1+\frac{b^2{y_i}\rho|u_{s_1,i}|^2}{8\sin^2\theta(1+b^2{y_i})}+\frac{b^2{y_i}\rho|u_{s_2,i}|^2}{8\sin^2\theta(1+b^2{y_i})}}\,\mathrm{d}{y_i}
\\&=\int_0^{\infty}\frac{b^2}{\gamma_i}\left(1+\frac{\frac1{b^2}-\frac1{\gamma_i}}{y_i+\frac1{\gamma_i}}\right)e^{-{y_i}}\,\mathrm{d}{y_i}
\\&=\frac{b^2}{\gamma_i}\left(1+\left(\frac1{b^2}-\frac1{\gamma_i}\right)J\left(\frac1{\gamma_i}\right)\right).
\end{split}
\end{equation}
where $\gamma_i=b^2+\alpha_i+\beta_i$,
$\alpha_i=\frac{\rho{b}^2|u_{s_1,i}|^2}{8\sin^2\theta}$, and
$\beta_i=\frac{\rho{b}^2|u_{s_2,i}|^2}{8\sin^2\theta}$. From
\cite{IEEEconf:8}, function $J(\nu)$ can be expressed as
\begin{equation}\label{equ:8}
\begin{split}
J(\nu)&=\int_0^{\infty}\frac{e^{-\mu}}{\mu+\nu}\,\mathrm{d}{\mu}
\\&=-e^{\nu}\left(\varphi+\ln{\nu}+\overset{\infty}{\underset{j=1}{\boldsymbol\sum}}\left((-1)^j\nu^j/(j!j)\right)\right)
\end{split}
\end{equation}
where $\varphi$ is the Euler constant. Then we assume $\rho$ is
large enough to work out the asymptotic solution. If
$\nu=c_1\rho^{-1}+O\left(c_2\rho^{-2}\right)\,(0<c_1,c_2<\infty)$,
when $\rho$ is large enough,
\begin{equation}\label{equ:9}
J(\nu)=\ln\rho+O(|\ln{c_1}|).
\end{equation}
Meanwhile, we have
\begin{equation}\label{equ:10}
\frac1{\gamma_i}=\frac1{b^2}\frac{8\sin^2\theta\rho^{-1}}{|u_{s_1,i}|^2+|u_{s_2,i}|^2}+O\left(\frac{\rho^{-2}}{|u_{s_1,i}|^2+|u_{s_2,i}|^2}\right).
\end{equation}
Then
\begin{equation}\label{equ:11}
\frac{b^2}{\gamma_i}\left(1+\left(\frac1{b^2}-\frac1{\gamma_i}\right)J\left(\frac1{\gamma_i}\right)\right)=\qquad\qquad\qquad\;
\end{equation}
\begin{equation*}
\frac{8\sin^2\theta{b^{-2}}\rho^{-1}}{|u_{s_1,i}|^2+|u_{s_2,i}|^2}\left(\ln\rho+O\left(\left|\ln\left(|u_{s_1,i}|^2+|u_{s_2,i}|^2\right)\right|\right)\right).
\end{equation*}
On the other hand, we denote
$K=\int_0^{\frac{\pi}{2}}\sin^{2(N+1)}\theta\,\mathrm{d}\theta$.
When $\rho$ is large enough, we can write
\begin{equation}\label{equ:14}
P_{PE,d_k}=\frac{Kb^{-2N}\rho^{-(N+1)}\ln^N\rho}{\overset{N}{\underset{j=1}{\boldsymbol\sum}}|u_{s_k,j}|^2{\overset{N}{\underset{j=1}{\boldsymbol\prod}}\left(|u_{s_k,j}|^2+|u_{s_{\bar{k}},j}|^2\right)}}.
\end{equation}
Since $\mathcal {E}(|u_{s_k,i}|^2)=\mathcal
{E}(|x_{s_k,i}-\hat{x}_{s_k,i}|^2)=4\kappa$,
\begin{equation}\label{equ:15}
\mathcal{E}(P_{sys})\sim\frac{\rho^{-(N+1)}\ln^N\rho}{\kappa\tau^N}.
\end{equation}
To minimize $\mathcal {E}(P_{sys})$, we should enlarge the value of
$\kappa\tau^N$. So the power allocation factors is worked out by
maximizing $\kappa\tau^N$ subject to the power constraint
$2\kappa+\tau=1$.~~~~~~~~~~~~~~~~~~{\ding{110}}
%=====================================================subsection B=============================================
\subsection{Instantaneous $s\rightarrow{r}$ CSI based Power Allocation}
Since the coefficients of $s\rightarrow{r}$ link should be notified
to destinations by relays for decoding, it is rational to assume
that the instantaneous $s\rightarrow{r}$ CSI can also be obtained by
sources without extra overheads. In this case, power is only
reallocated between the two sources while remains unchanged for the
relays. Then we get the instantaneous $s\rightarrow{r}$ CSI based
power allocation scheme by water filling \cite{IEEEconf:10} as
follows.
\begin{theorem}\label{theo2}
The instantaneous $s\rightarrow{r}$ CSI based optimal power
allocation scheme is
\begin{equation}\label{equ:16}
\begin{cases}
\kappa_{k,i}={\frac{2\kappa_i|g_{\bar{k},i}|^2}{|g_{k,i}|^2+|g_{{\bar{k}},i}|^2}},
\;\kappa_{{\bar{k}},i}={\frac{2\kappa_i|g_{k,i}|^2}{|g_{k,i}|^2+|g_{{\bar{k}},i}|^2}},
\\\kappa_i=\left(\frac{4N\kappa}{W}+\frac{\sum_{j=1}^{W}|g_j|^{-1}}{W\rho}-\frac1{|g_i|^2\rho}\right)^+,
\\\sum_{j=1}^{N}\kappa_j=2N\kappa,
\\\kappa=\frac1{2(N+1)},\;\tau=\frac{N}{N+1}
\end{cases}
\end{equation}
where
$|g_i|^2=\frac{|g_{k,i}g_{\bar{k},i}|^2}{|g_{k,i}|^2+|g_{\bar{k},i}|^2}$,
$\kappa_{s_k,i}$ is the power allocation factor of the $k$-th source
when transmitting to the $i$-th relay, and $W$ is the number of
positive $\kappa_i$. Moreover, instantaneous $s\rightarrow{r}$ CSI
is also used to pre-equalize the channels phase which produces a
coherent superposition of signals from two sources which can achieve
better performance.~~~~~~~~~~~~~~~~~~~~{\ding{111}}
\end{theorem}

\emph{Proof:} When $s\rightarrow{r}$ CSI is available at sources,
power is only reallocated between two sources. So the $\kappa$ and
$\tau$ is unchanged. In the $(2i-1)$-th time slot, the $i$-th relay
receives the mixed signals from the two sources which can be seen as
a multi-access model. We then turn to the achievable capacity region
of the channel between each source and the $i$-th relay when joint
ML decoding is performed. As well known, the mutual information
between the sources and relay on the channel realization
$\boldsymbol{g}_i$ is
\begin{equation}\label{equ:17}
I(s_k,s_{\bar{k}};r_i|\boldsymbol{g}_i)=\frac1{2}\log\big(1+2(|g_{k,i}|^2\kappa_{k,i}+|g_{{\bar{k}},i}|^2\kappa_{{\bar{k}},i})\rho\big)
\end{equation}

According to \cite{IEEEconf:9}, we suppose that each source splits
its power into the same pieces, i.e.,
$2\kappa_{k,i}\rho=M\triangle{\rho}_k$ for the $k$-th source. Two
sources alternatively pour one piece of their power into the
channels to gain the rate growth $\triangle{R}(s_{k}^m)$ in the
$m$-th round. Let $\triangle{\rho}_{k}\rightarrow{0}$, then
\begin{equation}\label{equ:18}
\begin{split}
&\triangle{R}(s_{k}^m)=\frac1{2}|g_{k,i}|^2\triangle{\rho_k}\eta_{i,m},\;
\\&\triangle{R}(s_{\bar{k}}^m)=\frac1{2}|g_{\bar{k},i}|^2\triangle{\rho_{\bar{k}}}\eta_{i,m},
\end{split}
\end{equation}
where
\begin{equation}
\eta_{i,m}=\frac{1}{1+m\underset{j=1}{\overset{2}{\sum}}{|g_{j,i}|^2\triangle\rho_{j}}}.
\end{equation}
Then the achievable capacity of the channel between the $k$-th
source and the $i$-th relay is
\begin{equation}\label{equ:19}
I(s_k;r_i|\boldsymbol{g}_i)=\int_0^{2\kappa_{k,i}\rho}\frac{\frac1{2}|g_{k,i}|^2}{1+m\underset{j=1}{\overset{2}{\sum}}{|g_{j,i}|^2\triangle\rho_{j}}}\,\mathrm{d}\rho_k.
\end{equation}
Since in joint ML decoding, $\kappa_{k,i}$ and
$\kappa_{{\bar{k}},i}$ are already known. So in (\ref{equ:20}), by
replacing $\triangle\rho_{\bar{k}}$ with
$\frac{\kappa_{\bar{k},i}}{\kappa_{k,i}}\triangle\rho_{k}$,
\begin{equation}\label{equ:20}
I(s_k;r_i|\boldsymbol{g}_i)=\int_0^{2\kappa_{k,i}\rho}\frac{\frac1{2}|g_{k,i}|^2\,\mathrm{d}\rho_k}{1+(|g_{k,i}|^2+\frac{\kappa_{\bar{k},i}}{\kappa_{k,i}}|g_{\bar{k},i}|^2)\rho_{k}}.
\end{equation}
Thus we get
\begin{equation}\label{equ:21}
I(s_k;r_i|\boldsymbol{g}_i)=\frac{\kappa_{k,i}|g_{k,i}|^2}{\kappa_{k,i}|g_{k,i}|^2+\kappa_{\bar{k},i}|g_{{\bar{k}},i}|^2}I(s_k,s_{\bar{k}};r_i|\boldsymbol{g}_i).
\end{equation}

Then power allocation scheme for the two sources in the $(2i-1)$-th
time slot, namely local power allocation when $s\rightarrow{r}$ CSI
is available, is to make
\begin{equation}\label{equ:22}
\max_{\kappa_{k,i},\kappa_{{\bar{k}},i}}\min\{I(s_k;r_{i}|\boldsymbol{g}_i),I(s_{\bar{k}};r_{i}|\boldsymbol{g}_i)\}.
\end{equation}
Due to the symmetrical quality of the channel model, we should keep
the balance of the system, i.e.,
$I(s_k;r_{i}|\boldsymbol{g}_i)=I(s_{\bar{k}};r_{i}|\boldsymbol{g}_i)$.
Then
\begin{equation}\label{equ:23}
{\kappa_{k,i}}|{g_{k,i}}|^2={\kappa_{{\bar{k}},i}}|{g_{{\bar{k}},i}}|^2,~
\kappa_{k,i}+\kappa_{{\bar{k}},i}=2\kappa.
\end{equation}
So we get the local power  allocation scheme as
\begin{equation}\label{equ:24}
\kappa_{k,i}={\frac{2\kappa|g_{\bar{k},i}|^2}{|g_{k,i}|^2+|g_{{\bar{k}},i}|^2}},\;
\kappa_{{\bar{k}},i}={\frac{2\kappa|g_{k,i}|^2}{|g_{k,i}|^2+|g_{{\bar{k}},i}|^2}}.
\end{equation}

After local power allocation, the channels from the two sources to
the $i$-th relay suffer the same fading
$\frac{|g_{k,i}g_{\bar{k},i}|^2}{|g_{k,i}|^2+|g_{\bar{k},i}|^2}$.
Then a frame period can be divided into $N$ orthogonal time division
channels with channel fading $|g_i|^2$ for the $i$-th channel.
According to \cite{IEEEconf:10}, we get the global power allocation
scheme (\ref{equ:16}) by water filling
scheme.~~~~~~~~~~~~~~~~~~~~~~~~~~~~~~~~~~~~{\ding{110}}
%========================================================================
\subsection{Precoders Design}
From (\ref{equ:14}), it is obvious that the system can not achieve
the full diversity gain, i.e., the denominator of (\ref{equ:14}) has
the chance to equal to $0$, which decreases the diversity orders.
Precoder is then applied to enhance the diversity gain.
\subsubsection{Isolated Precoder} If only statistical CSI is
available, precoders are designed in the same way for both sources,
i.e., \emph{isolated precoder}. We follow the precoder design in
MIMO \cite{IEEEconf:6} which achieves full diversity gain. Then for
each source, the precoder matrix is
\begin{equation}\label{equ:25}
\boldsymbol\Theta_s=\frac1{\sqrt{N}}
\begin{pmatrix}
1 & \alpha_1 & \cdots & \alpha_1^{N-1}\\
\vdots & \vdots & & \vdots\\
1 & \alpha_N & \cdots & \alpha_N^{N-1}
\end{pmatrix}_{N\times{N}},
\end{equation}
where $\{\alpha_i\}_{i=1}^N$ have unit modulus. Thus, the
transmitted signals for the $k$-th source becomes
$\sqrt{\kappa}\boldsymbol{\Theta}_{s}\textbf{x}_{s_k}$.
\begin{remark}\label{rmk1}
When $\rho\rightarrow\infty$, by applying statistical CSI based
power allocation scheme and isolated precoders in each source, the
average PEP of $d_k$ is then
\begin{equation}\label{equ:26}
P_{PE,d_k}=\frac{{(N+1)^{N+1}}K\rho^{-(N+1)}\ln^N\rho}{{(N/2)^N}\overset{N}{\underset{j=1}{\boldsymbol\sum}}\mu_{s_k,j}{\overset{N}{\underset{j=1}{\boldsymbol\prod}}\left(\mu_{s_1,j}+\mu_{s_2,j}\right)}},
\end{equation}
where
$\mu_{s_k,j}=\frac1{{N}}\left|{\boldsymbol\sum_{n=1}^N}\alpha_i^{n-1}{u}_{s_k,n}\right|^2$.~~~~~~~~~~~~~~~~~~~~~~~~~~~~~~~~~~~~~~~~{\ding{111}}
\end{remark}

By applying precoder, the denominator of equation (\ref{equ:26})
equals to $0$ if and only if the frame can be successfully decoded.
Then the whole system can achieve full diversity gain. However,
isolated precoders design is not suitable for instantaneous
$s\rightarrow{r}$ CSI based power allocation scheme, in which it can
not achieve the full diversity gain. More specifically, for the
$i$-th relay,
$\mu_{s_k,i}+\mu_{s_{\bar{k}},i}=\frac1{{N}}\left|{\boldsymbol\sum_{j=1}^N}\alpha_i^{j-1}({u}_{s_k,j}+{u}_{s_{\bar{k}},j})\right|^2$.
A wrong decoding of the $j$-th symbol $\hat{x}_{s_k,j}$ and
$\hat{x}_{s_{\bar{k}},j}$ in the two symbol vectors may cause
${u}_{s_k,j}+{u}_{s_{\bar{k}},j}=0$, which lead to zero in the
denominator of PEP expression and hence the lower diversity gain.
Then we propose the \emph{distributed precoder}.
\subsubsection{Distributed Precoder}We first construct a $2N\times{2N}$ matrix as (\ref{equ:25}). Then
arbitrary $N$ rows is selected to form a new matrix
$\boldsymbol\Theta$. The precoder matrix $\boldsymbol{\Theta}_{s_1}$
for $s_1$ comes from the odd columns of the $\boldsymbol\Theta$ and
the precoder matrix $\boldsymbol{\Theta}_{s_2}$ for $s_2$ comes from
the even columns of the $\boldsymbol\Theta$, i.e., for the $k$-th
source,
\begin{equation}\label{equ:27}
\begin{split}
&\boldsymbol\Theta_{s_k}=\frac1{\sqrt{N}}
\begin{pmatrix}
\alpha_1^{k-1} & \cdots\ & \alpha_1^{2j+k-1} & \cdots & \alpha_1^{2N+k-3}\\
\vdots & & \vdots & & \vdots\\
\alpha_N^{k-1} & \cdots\ & \alpha_N^{2j+k-1} & \cdots &
\alpha_N^{2N+k-3}
\end{pmatrix}_{N\times{N}}
\end{split}
\end{equation}
where $j\in\{0,\cdots,N-1\}$. By this mean, signal superposed in the
$i$-th relay will be equal to
$\frac{{\kappa_i}|g_{k,i}g_{{\bar{k}},i}|^2}{|g_{k,i}|^2+|g_{{\bar{k}},i}|^2}\boldsymbol\theta^i\textbf{x}_{s}$
where $\boldsymbol\theta^i$ denotes the $i$-th row of
$\boldsymbol\Theta$. It means that in PEP expression,
$\mu_{s_k,i}+\mu_{s_{\bar{k}},i}=\frac1{{N}}\left|{\boldsymbol\sum_{j=1}^N}\alpha_i^{j-1}({u}_{s_k,j}+\alpha_i{u}_{s_{\bar{k}},j})\right|^2$,
which achieves full diversity gain.
%The right hand side of (\ref{equ:12}) is deduced under the
%assumption that symbols in $s\rightarrow{d}$ link are not multiplied
%by $\textbf{\textrtailn}_{s_k}$, while in the real multicast system
%with DP, both $s\rightarrow{d}$ and $s\rightarrow{r}$ link are
%effected by the $\textbf{\textrtailn}_{s_k}$ which can bring more
%diversity gain in $s\rightarrow{d}$ link, thus get lower SFEP.
%=======================================================section4=====================================================================
\section{Numerical Results}\label{sec:4}
\begin{figure}[!t]
\centering
\includegraphics[width=3.5in,angle=0]{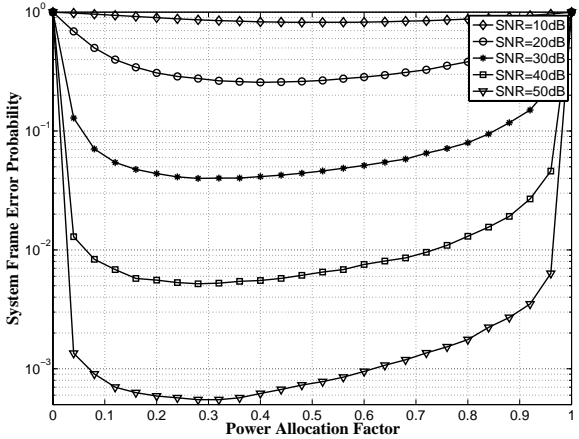}
\caption{SFEP in two-relay scenario with different power allocation
schemes using 4-QAM modulation. Power allocation factor is the value
of $2\kappa$.} \label{fig2}
\end{figure}

In our Monte-Carlo simulations, we choose $4$-QAM modulation with
$2$ relays. Each frame has $4$ symbols and each SFEP value is
simulated by $10^6$ $i.i.d$ frames. Fig.~\ref{fig2} shows the SFEP
under statistical CSI based power allocation schemes with different
values of power allocation factor $\kappa$. We can see that when
$\rho$ is large enough, systems reaches the lowest SFEP for
$2\kappa=1/3$ and $\tau=2/3$, which validates \emph{Theorem~1}.
\begin{figure}[!t]
\centering
\includegraphics[width=3.5in,angle=0]{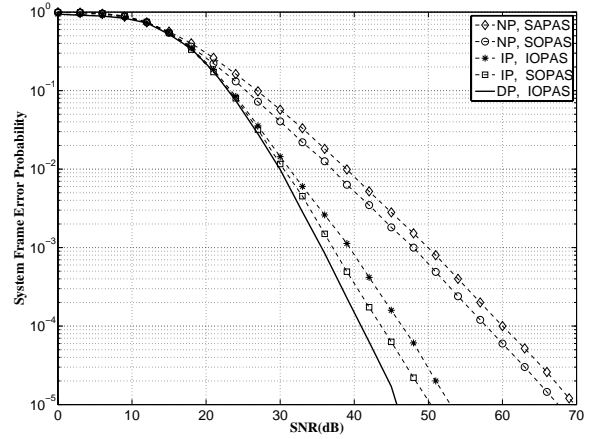}
\caption{SFEP in two-relay scenario using 4-QAM modulation in 5
cases.} \label{fig3}
\end{figure}

Fig.~\ref{fig3} compares the SFEP in six cases: 1) Statistical CSI
based average power allocation scheme (SAPAS) without precoder (NP);
2) Statistical CSI based optimal power allocations scheme (SOPAS)
without precoder; 3) Instantaneous $s\rightarrow{r}$ CSI based
optimal power allocation scheme (IOPAS) with isolated precoders
(IP); 4) SOPAS with IP; 5) IOPAS with DP. By SAPAS, both sources and
relays consume the same power. We can see that power allocation
scheme and precoders can distinctly improve the performance. Since
IP with IOPAS can not achieve full diversity gain, it leads to worse
performance even than that with SOPAS. However, DP with IOPAS
achieves the best performance since it make the system achieve full
diversity with larger Euclidian distance.

\section{Conclusion}
In this letter, we propose two power allocation schemes and two
precoders for $2-N-2$ multicast systems with non-regenerative
network coding. Although IP can help SOPAS achieve full diversity
gain, it can not well help IOPAS. However, DP jointly with IOPAS can
make the system not only achieve full diversity gain, but also
outperform any other combinations of power allocation schemes and
precoders in the term of SFEP.

% trigger a \newpage just before the given reference
% number - used to balance the columns on the last page
% adjust value as needed - may need to be readjusted if
% the document is modified later
%\IEEEtriggeratref{8}
% The "triggered" command can be changed if desired:
%\IEEEtriggercmd{\enlargethispage{-5in}}

% references section

% can use a bibliography generated by BibTeX as a .bbl file
% BibTeX documentation can be easily obtained at:
% http://www.ctan.org/tex-archive/biblio/bibtex/contrib/doc/
% The IEEEtran BibTeX style support page is at:
% http://www.michaelshell.org/tex/ieeetran/bibtex/
%\bibliographystyle{IEEEtran}
% argument is your BibTeX string definitions and bibliography database(s)
%\bibliography{IEEEabrv,../bib/paper}
%
% <OR> manually copy in the resultant .bbl file
% set second argument of \begin to the number of references
% (used to reserve space for the reference number labels box)

% that's all folks

\end{document}